\documentclass[a4paper,fleqn,usenatbib]{mnras}

\usepackage{newtxtext,newtxmath}

\usepackage[T1]{fontenc}
\usepackage{ae,aecompl}

\usepackage{graphicx}	
\usepackage{amsmath}	
\usepackage{amssymb}	

\title[A new planetary mass companion around TYC~8998-760-1]{The Young Suns Exoplanet Survey: Detection of a wide orbit planetary mass companion to a solar-type Sco-Cen member\thanks{Based on observations collected at the European Organisation for Astronomical Research in the Southern Hemisphere under ESO programs 099.C-0698(A), 0103.C-0371(A), and 2103.C-5012(A,B).}}

\author[A.~J.~Bohn et al.]{
A.~J.~Bohn,$^{1}$\thanks{E-mail: bohn@strw.leidenuniv.nl}
M.~A.~Kenworthy,$^{1}$
C.~Ginski,$^{2}$
C.~F.~Manara,$^{3}$
M.~J.~Pecaut,$^{4}$
\newauthor
J.~de~Boer,$^{1}$
C.~U.~Keller,$^{1}$
E.~E.~Mamajek,$^{5,6}$
T.~Meshkat,$^{7}$
M.~Reggiani,$^{8}$
\newauthor
K.~O.~Todorov,$^{2}$
and F.~Snik$^{1}$
\\
$^{1}$Leiden Observatory, Leiden University, PO Box 9513, 2300 RA Leiden, The Netherlands\\
$^{2}$Sterrenkundig Instituut Anton Pannekoek, Science Park 904, 1098 XH Amsterdam, The Netherlands\\
$^{3}$European Southern Observatory, Karl-Schwarzschild-Strasse 2, 85748 Garching bei M\"unchen, Germany\\
$^{4}$Rockhurst University, Department of Physics, 1100 Rockhurst Road, Kansas City, MO 64110, USA\\
$^{5}$Jet Propulsion Laboratory, California Institute of Technology, 4800 Oak Grove Drive, M/S 321-100, Pasadena, CA, 91109, USA\\
$^{6}$Department of Physics \& Astronomy, University of Rochester, Rochester, NY 14627, USA\\
$^{7}$IPAC, California Institute of Technology, M/C 100-22, 1200 East California Boulevard, Pasadena, CA 91125, USA\\
$^{8}$Institute of Astronomy, KU Leuven, Celestijnenlaan 200D, B-3001 Leuven, Belgium
}

\date{Accepted 2019 December 5. Received 2019 November 4; in original form 2019 September 6.}

\pubyear{2019}

\begin{document}
\label{firstpage}
\pagerange{\pageref{firstpage}--\pageref{lastpage}}
\maketitle

\begin{abstract}
The Young Suns Exoplanet Survey (YSES) consists of a homogeneous sample of 70 young, solar-mass stars located in the Lower Centaurus-Crux subgroup of the Scorpius-Centaurus association with an average age of $15\pm3$\,Myr.
We report the detection of a co-moving companion around the K3IV star TYC~8998-760-1 (2MASSJ13251211-6456207) that is located at a distance of $94.6\pm0.3$\,pc using SPHERE/IRDIS on the VLT.
Spectroscopic observations with VLT/X-SHOOTER constrain the mass of the star to  $1.00\pm0.02\,M_{\sun}$ and an age of $16.7\pm1.4\,$Myr.
The companion TYC~8998-760-1~b is detected at a projected separation of 1.71\arcsec, which implies a projected physical separation of 162\,au.
Photometric measurements ranging from $Y$ to $M$ band provide a mass estimate of $14\pm3\,M_\mathrm{jup}$ by comparison to BT-Settl and AMES-dusty isochrones, corresponding to a mass ratio of $q=0.013\pm0.003$ with respect to the primary.
We rule out additional companions to TYC~8998-760-1 that are more massive than $12\,M_\mathrm{jup}$ and farther than 12\,au away from the host.
Future polarimetric and spectroscopic observations of this system with ground and space based observatories will facilitate testing of formation and evolution scenarios shaping the architecture of the circumstellar environment around this 'young Sun'.
\end{abstract}




\section{Introduction}
\label{sec:introduction}

With the advent of extreme adaptive optics (AO) assisted, high-contrast imaging instruments at the current generation of 8-m class telescopes, the search and characterisation of directly imaged extra-solar planets has gained momentum.
The large scale guaranteed time observing campaigns that are currently carried out with these instruments such as the Gemini Planet Imager Exoplanet Survey \citep[GPIES;][]{Macintosh2014} or the SpHere INfrared survey for Exoplanets \citep[SHINE;][]{Chauvin2017b}, can constrain the occurrence rates of gas giant companions in wide orbits \citep{Nielsen2019}.
In addition to these ongoing statistical evaluations, both surveys have already produced many high-impact results by new detections of giant companions \citep[e.g.][]{Macintosh2015,Chauvin2017b,Keppler2018} as well as spectral and orbital characterisations of established members among almost twenty directly imaged extra-solar planets \citep[e.g.][]{Galicher2014,Wang2016,Wang2018,Greenbaum2018,Samland2017,Chauvin2018,Muller2018,Cheetham2019,Lagrange2019}.

Most of these directly imaged companions, however, are detected around stars that are more massive than the Sun.
To obtain a statistically significant estimate on the occurrence rates of giant sub-stellar companions on wide orbits around solar-type stars, we started the Young Suns Exoplanet Survey (YSES; Bohn et al. in prep.).
YSES targets a homogeneous sample of 70 young, solar-type stars located in the Lower-Centaurus Crux subgroup of the Scorpius-Centaurus association \citep[Sco-Cen;][]{deZeeuw1999}.
Based on common kinematics and activity signatures, all YSES targets have been confirmed by \citet{Pecaut2016} as members of the LCC;
Gaia DR2 parallaxes and proper motions corroborate this membership status \citep{Gaia2018}.
In addition to the small range of stellar masses, the YSES targets are homogeneous in terms of stellar ages and distances.
This enables self-consistent reference star differential imaging \citep[RDI;][]{Smith1984,Lafreniere2007} to increase the contrast performance at close separations (Bohn et al. in prep.) and minimises uncertainties on the properties of identified companions due to poorly constrained system ages.

One object within our sample is TYC~8998-760-1 (2MASSJ13251211-6456207) at a distance of $94.6\pm0.3$\,pc \citep{Bailer-Jones2018,Gaia2018}.
Based on new observations of the system we revised the main stellar properties (Section~\ref{subsec:results_stellar_properties}) as summarised in Table \ref{tbl:stellar_properties}.
\begin{table}
\caption{Stellar properties of TYC~8998-760-1.}
\label{tbl:stellar_properties}
\def\arraystretch{1.2}
\setlength{\tabcolsep}{8pt}
\begin{tabular}{@{}lll@{}}
\hline
Parameter & Value & Reference(s)\\ 
\hline
Main identifier & TYC~8998-760-1 & (1)\\
2MASS identifier & J13251211-6456207 & (2)\\
Right Ascension (J2000) & 13:25:12.13 & (3) \\
Declination (J2000) & -64:56:20.69 & (3) \\
Spectral Type &  K3IV & (4,5) \\
Mass [$M_\mathrm{\sun}$] & $1.00\pm0.02$ & (5) \\
$T_\mathrm{eff}$ [K] & $4573\pm10$ & (5)\\
$\log\left(L/L_\mathrm{\sun}\right)$ [dex] & $-0.339\pm0.016$ & (5)\\
Age [Myr] & $16.7\pm1.4$ & (5)\\
Parallax [mas] & $10.540\pm0.031$ & (3) \\
Distance [pc] & $94.6\pm0.3$ & (6) \\
Proper motion (RA) [mas / yr] & $-40.898\pm0.045$ & (3) \\
Proper motion (Dec) [mas / yr] & $-17.788\pm0.043$ & (3) \\
$B$ [mag]& $11.94$ & (7) \\
$V$ [mag] & $11.13$ & (7) \\
$R$ [mag] & $10.61$ & (7) \\
$J$ [mag] & $9.07$ & (2) \\
$H$ [mag] & $8.56$ & (2) \\
$K_\text{s}$ [mag] & $8.39$ & (2) \\
$W1$ [mag] & $8.37$ & (8) \\
$W2$ [mag] & $8.38$ & (8) \\
$W3$ [mag] & $8.32$ & (8) \\
$W4$ [mag] & $>8.43$ & (8) \\
\hline
\end{tabular}
\textbf{References.} (1)~\citet{Hog2000}; (2)~\citet{Cutri2012a}; (3)~\citet{Gaia2018}; (4)~\citet{Pecaut2016}; (5)~Section~\ref{subsec:results_stellar_properties} of this work; (6)~\citet{Bailer-Jones2018}; (7)~\citet{Zacharias2005}; (8)~\citet{Cutri2012b}
\end{table}

In Section~\ref{sec:observations} of this article we describe the observations that we carried out on TYC~8998-760-1 and in Section~\ref{sec:data_reduction} we explain our data reduction strategies.
In Section~\ref{sec:results} we illustrate how we detect a co-moving planetary mass companion around TYC~8998-760-1 and in Section~\ref{sec:discussion} we discuss the derived properties of this companion.
The conclusions of the article are presented in Section~\ref{sec:conclusions}.

\section{Observations}
\label{sec:observations}

Our observations of the system can be classified by two categories:
(i) medium-resolution spectrographic observations of the host with VLT/X-SHOOTER and (ii) high-contrast imaging data collected with VLT/SPHERE and VLT/NACO.
Whereas the former data aims for a precise characterisation of the host star, the latter observations facilitate an accurate astrometric and photometric characterisation of the companion around TYC~8998-760-1.

\subsection{X-SHOOTER}
\label{subsec:observations_XSHOOTER}

We observed TYC~8998-760-1 with X-SHOOTER \citep{vernet11} on the night of May 23, 2019, in excellent atmospheric conditions with an average seeing of 0\farcs54 (PI: A.~Bohn; ESO ID: 2103.C-5012(A)).
X-SHOOTER was operated in SLT mode providing medium resolution spectra from $300-2500$\,nm.
We chose slit widths of 0\farcs8, 0\farcs4, and 0\farcs4 with corresponding exposure times of 210\,s, 120\,s, and $3\times80$\,s for UVB, VIS, and NIR\footnote{The individual integration time for the NIR arm was 80\,s and each exposure is composed of 3 sub-integrations (NDIT).} subsystems, respectively.
Applying two nodding cycles along the slit for background subtraction at NIR wavelengths, yielded total integration times of 840\,s, 480\,s, and 960\,s for the three subsystems.
For flux calibration we took additional spectra with a wide slit configuration of 5\arcsec and exposure times of 15\,s, 60\,s and $4\times15$\,s for UVB, VIS, and NIR arm, respectively.

\subsection{SPHERE}
\label{subsec:observations_SPHERE}

\begin{table*}
\caption{High-contrast observations of TYC~8998-760-1.
}
\label{tbl:observations}
\def\arraystretch{1.2}
\setlength{\tabcolsep}{10pt}
\begin{tabular}{@{}llllllllll@{}}
\hline\hline
Observation date & Instrument & Mode & Filter & FWHM & NEXP$\times$NDIT$\times$DIT & $\Delta\pi$ & $\langle\omega\rangle$ & $\langle X\rangle$ & $\langle\tau_0\rangle$ \\
(yyyy-mm-dd) & & & & (mas) & (1$\times$1$\times$s) & (\degr) & (\arcsec) & & (ms)\\
\hline
2017-07-05 & SPHERE & CI & $J$ & 46.7 & 4$\times$2$\times$32 & 1.11 & 1.12 & 1.54 & 3.15\\
2017-07-05 & SPHERE & CI & $H$ & 52.3 & 4$\times$1$\times$32 & 0.50 & 1.22 & 1.52 & 2.90 \\
2019-03-17 & SPHERE & CI & $K_s$ & 64.2 & 6$\times$2$\times$32 & 2.26 & 1.11 &  1.31 & 3.15 \\
2019-03-23 & SPHERE & DBI & $Y23$ & 37.2 / 37.9 & 4$\times$3$\times$64 & 3.84 & 0.41 & 1.38 & 9.30 \\
2019-03-23 & SPHERE & DBI & $J23$ & 40.1 / 41.8 & 4$\times$3$\times$64 & 3.72 & 0.40 & 1.41 & 10.75\\
2019-03-23 & SPHERE & DBI & $H23$ & 47.5 / 49.5 & 4$\times$3$\times$64 & 3.60& 0.43 & 1.44 & 10.83\\
2019-03-23 & SPHERE & DBI & $K12$ & 60.2 / 63.6 & 4$\times$3$\times$64 & 3.45 & 0.53 & 1.49 & 8.75\\
2019-05-18 & NACO & CI & $L'$ & 125.0 & 30$\times$600$\times$0.2 & 22.99 & 0.88 & 1.32 & 2.32\\
2019-06-03 & NACO & CI & $M'$ & 131.6 & 112$\times$900$\times$0.045 & 50.15 & 0.78 & 1.33 & 3.69\\
\hline
\end{tabular}\\
\textbf{Notes.} The applied mode is either classical imaging (CI) with a broadband filter or dual-band imaging (DBI) with two intermediate band filters simultaneously.
FWHM denotes the full width at half maximum that we measure from the average of the non-coronagraphic flux images that are collected for each filter. For NACO data these are equivalent to the science exposures of the star.
NEXP describes the number of exposures, NDIT is the number of sub-integrations per exposure and DIT is the detector integration time of an individual sub-integration.
$\Delta\pi$ denotes the amount of parallactic rotation during the observation and $\langle\omega\rangle$, $\langle X\rangle$, and $\langle\tau_0\rangle$ represent the average seeing, airmass, and coherence time, respectively.
\end{table*}

The first part of our high-contrast imaging observations were carried out with SPHERE \citep{Beuzit2019}, mounted at the Naysmith platform of Unit 3 telescope (UT3) at ESO's VLT.
SPHERE is assisted by the SAXO extreme AO system \citep{Fusco2006} to deliver diffraction limited imaging data.
We used the infrared dual-band imager and spectrograph \citep[IRDIS;][]{Dohlen2008} in classical imaging (CI) and dual-band imaging \citep[DBI;][]{Vigan2010} modes.
To block the stellar flux and to enable longer exposure times we used SPHERE's apodized Lyot coronagraph \citep{Soummer2005}.
We obtained additional center frames by applying a sinusoidal pattern to the instrument's deformable mirror to determine the position of the star behind the coronagraph.
This creates four waffle spots around the star that can be used for precise centration\footnote{See description in the latest version of the SPHERE manual: \url{https://www.eso.org/sci/facilities/paranal/instruments/sphere/doc.html}}. 
For photometric calibration we took additional flux images by offsetting the stellar point spread function (PSF) from the coronagraphic mask and used a neutral density filter to avoid saturation of the detector.
All observations were carried out in pupil tracking mode to enable post-processing based on RDI within the scope of the survey (Bohn et al. in prep.).

We took short first epoch observations (Night: July 5, 2017; PI: M.~Kenworthy; ESO ID: 099.C-0698(A)) applying a broadband filter in \textit{J} and \textit{H} band\footnote{All filter profiles can be found at \url{https://www.eso.org/sci/facilities/paranal/instruments/sphere/inst/filters.html}}.
For second epoch observations (Night: March 17, 2019; PI: A.~Bohn; ESO ID: 0103.C-0371(A)), we scheduled a long sequence using the instrument's integral field spectrograph \citep[IFS;][]{Claudi2008} in extended mode in combination with IRDIS/CI in $K_s$ band.
The IFS provides low resolution spectra with a resolving power of $R=30$ ranging from $Y$ to $H$ band for the innermost field of view (1\farcs73$\times$1\farcs73) around the star.
Due to degrading weather conditions the observation was terminated after 384\,s.
In this aborted sequence, however, we detected a co-moving companion that was located outside the IFS's field of view.
We thus rearranged the observational setup aiming for optimal photometric characterisation of this companion.
These second epoch observations were obtained on the night of March 23, 2019, integrating for 768\,s with each of the $Y23$, $J23$, $H23$, and $K12$ DBI filter combinations.
A detailed description of the observations, applied filters, and weather conditions is presented in Table~\ref{tbl:observations}.

\subsection{NACO}
\label{subsec:observations_NACO}

To constrain the thermal infrared spectral energy distribution (SED) of the companion, we took additional $L'$ and $M'$ band data (PI: A.~Bohn; ESO ID: 2103.C-5012(B)) with VLT/NACO \citep{Lenzen2003,Rousset2003}.
A summary of the observational parameters is presented in Table~\ref{tbl:observations}.
The instrument was operated in pupil-stabilised imaging mode and the detector readout was performed in cube mode to store each individual sub-integration.
As the star is faint at the observed wavelengths, no coronagraph was used.
We chose integrations times of 0.2\,s and 0.045\,s for the observations in $L'$ and $M'$ band, respectively, resulting in 3600\,s and 4536\,s total time on target.
In both configurations the science frames are unsaturated and the individual pixel counts are in the linear regime of the detector, so no additional flux calibration frames were required.

\section{Data reduction}
\label{sec:data_reduction}

\begin{figure*}
\centering
\includegraphics[width=\textwidth]{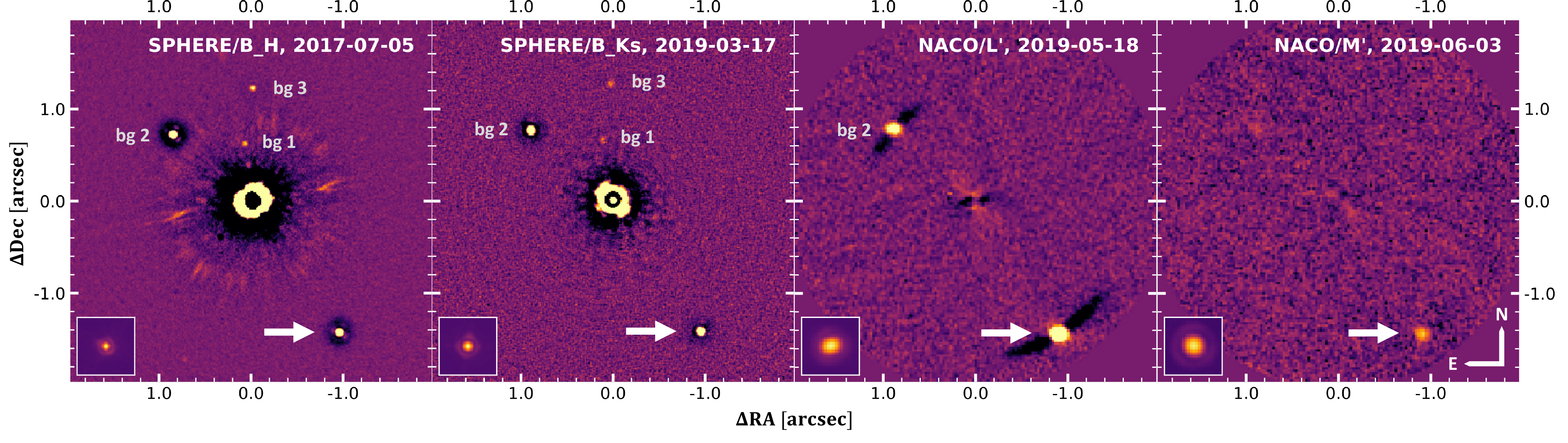}
\caption{
Reduced imaging data on TYC~8998-760-1.
We present four different epochs on the target that were collected in $H$, $K_s$, $L'$, and $M'$ band, respectively.
For the SPHERE data, an unsharp mask is applied;
the NACO results are reduced with ADI and the main principal component subtracted.
All images are presented with an arbitrary logarithmic colour scale to highlight off-axis point sources.
Proper motion analysis proves that all objects north of the star are background (bg) contaminants, while the object south-west of TYC~8998-760-1 (highlighted by the white arrow) is co-moving with its host.
This claim is supported by the very red colour of this object compared to the other point sources in the field.
In the lower left of the each figure we present the reduced non-coronagraphic flux image at the same spatial scale and field orientation.
For all images north points up and east towards the left.}
\label{fig:yses_companion}
\end{figure*}

\subsection{X-SHOOTER data}
\label{subsec:data_reduction_XSHOOTER}

The X-SHOOTER data were reduced using the ESO pipeline \citep{modigliani10} v3.2.0 run through the Reflex workflow. 
The pipeline includes bias and flat-field correction, wavelength calibration, spectrum rectification, flux calibration using a standard star observed in the same night, and spectrum extraction. 
As described in Section~\ref{sec:observations}, the target was observed with a set of wide slits of 5\arcsec, which have no slit losses, and another set of narrower slits providing higher spectral resolution. 
After the standard pipeline flux calibration, the data obtained with the wider slits shows good agreement in the flux between the three arms. 
The spectra obtained with the narrower slits show a lower flux than the ones with the wide slits by a factor $\sim$1.7, 2.7, and 2.5 in the three arms, respectively. 
The narrower slit spectra were adjusted in flux by this ratio in the UVB and NIR arms, and by a wavelength dependent ratio in the VIS arm to match the wide slit spectra. 
This final flux calibrated spectrum is in good agreement with previous non-simultaneous photometry.
The spectra were corrected for telluric absorption using the MOLECFIT tool \citep{molecfit1,molecfit2}.

\subsection{SPHERE data}
\label{subsec:data_reduction_SPHERE}

The SPHERE data were reduced with a custom processing pipeline based on the latest version of the PynPoint package \citep[version~0.8.1;][]{Stolker2019}.
This includes flatfielding, sky subtraction, and bad pixel correction by replacing bad pixels with the average value in a 5$\times$5 pixels sized box around the corresponding location.
We corrected for the instrumental anamorphic distortion in $y$ direction according to the description in the SPHERE manual.
For the data obtained in CI mode, we averaged both detector PSFs per exposure to minimise the effect of bad pixels.
Since the companion is not contaminated by stellar flux, we did not perform any advanced PSF subtraction.
We simply derotated the individual frames according to the parallactic rotation of the field and the static instrumental offset angle of 135\fdg99 required for correct alignment of pupil and Lyot stop, and we used the standard astrometric solution for IRDIS \citep{Maire2016}.
This provides a general true north correction of $-1\fdg75\pm0\fdg08$ and plate scales in the range of 12.283$\pm$0.01\,mas per pixel and $12.250\pm0.01$\,mas per pixel depending on the applied filter.

\subsection{NACO data}
\label{subsec:data_reduction_NACO}

For reduction of the NACO data, we used the same framework as applied for SPHERE including flatfielding, dark subtraction, and bad pixel correction.
There is a high readout noise that decreases exponentially throughout the cube, so we removed the first 5 frames of each cube.
The background subtraction was performed by an approach based on principal component analysis (PCA) as described in \citet{Hunziker2018} making use of the three distinct dither positions on the detector.
We masked a region of 0\farcs55 around the star and fitted 60 principal components to model sky and instrumental background.
After subtraction of this model, we aligned the stellar PSFs by applying a cross-correlation in the Fourier domain \citep{Guizar2008} and centred the aligned images by fitting a two-dimensional Gaussian function to the average of the stack.
Frame selection algorithms then reject all frames which deviate by more than 2$\sigma$ from the median flux within (i) a background annulus with inner and outer radii of 1\farcs6 and 1\farcs9 and (ii) an aperture with the size of the average PSF FWHM, resulting in 10.45\% and 10.05\% of our $L'$ and $M'$ band data being removed from the subsequent analysis.
All frames were derotated according to their parallactic angle and median combined.
As we have a sufficient amount of parallactic rotation for both datasets, we tested angular differential imaging \citep[ADI;][]{Marois2006} techniques for further analysis steps as described in the following Section.
For astrometric calibration of the results we adapted a plate scale of $27.20\pm0.06$\,mas per pixel and a true north correction of $0\fdg486\pm0\fdg180$ according to \citet{MussoBarcucci2019} and Launhardt et al. (in prep.).

\section{Results and analysis}
\label{sec:results}

Our first epoch observation with SPHERE reveals 16 off-axis point sources around TYC~8998-760-1 within the IRDIS field of view (11\farcs0$\times$12\farcs5).
We present the innermost 2\arcsec$\times$2\arcsec for several epochs and wavelengths in Figure~\ref{fig:yses_companion}.
All point sources in the field of view are consistent with background sources at 5$\sigma$ significance with the exception of the point source south-west of the star (highlighted by the white arrow) which has a proper motion consistent with being a co-moving companion (see analysis in Section~\ref{subsubsec:astrometric_analysis}).
This hypothesis is strongly supported by the very red colour of this object in comparison to the other sources in the field of view in Figure~\ref{fig:yses_companion}.
In order to constrain the properties of this companion, the properties of the host star - especially its age - need to be determined first.

\subsection{Stellar properties}
\label{subsec:results_stellar_properties}

\begin{figure}
\centering
\includegraphics[width=\columnwidth]{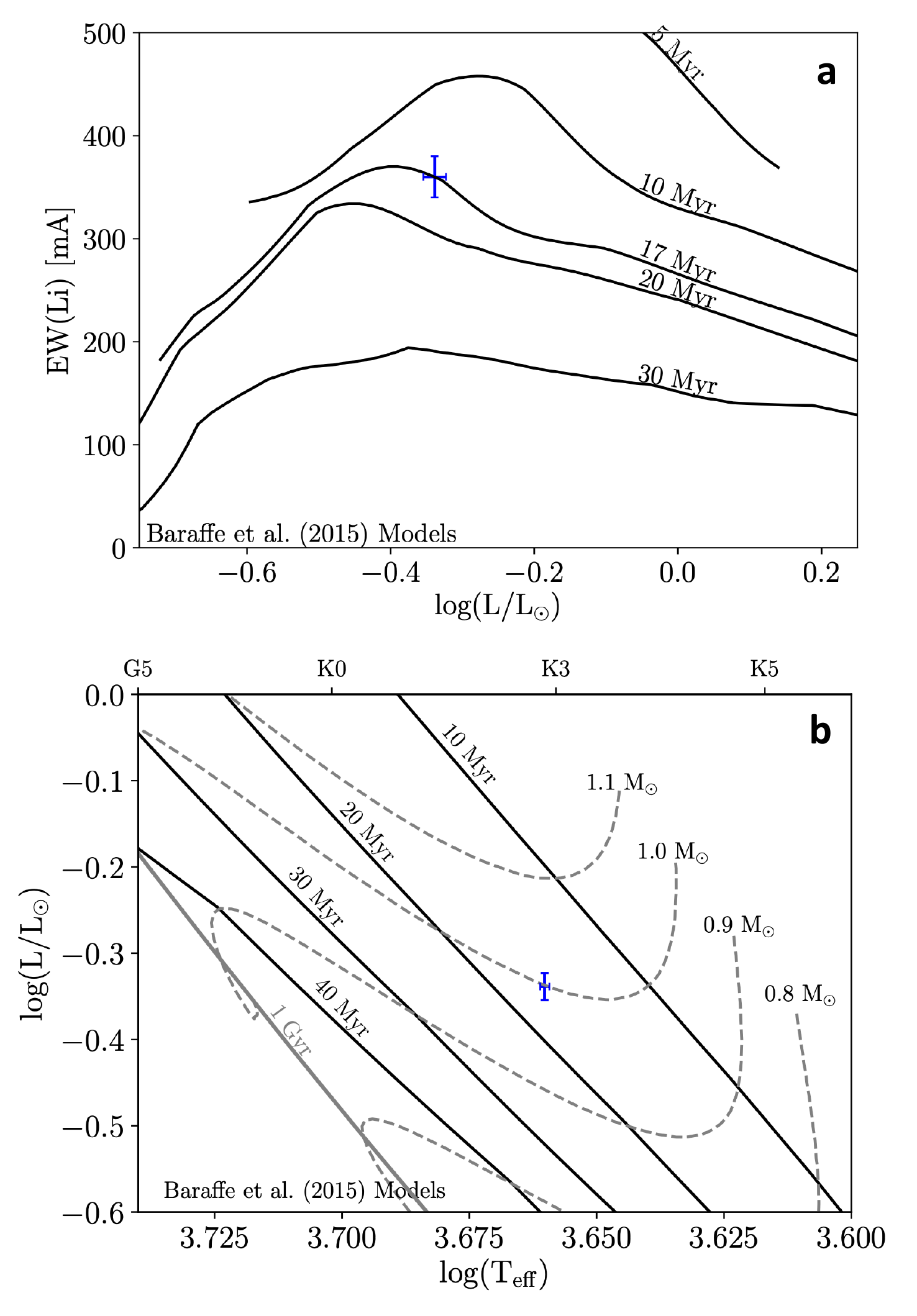}
\caption{
Stellar properties of TYC~8998-760-1.
\emph{Panel~(a)}: 
\citet{Baraffe2015} isochrones plotted for the Lithium-absorption equivalent width that we measure in the X-SHOOTER spectrum.
\emph{Panel~(b)}:
Hertzsprung-Russell diagram using the effective temperature that is  constrained by fitting BT-Settl models to Tycho-2, APASS, Gaia, 2MASS, and WISE photometry.
The isochronal tracks from \citet{Baraffe2015} are used to determine the stellar mass and age.
}
\label{fig:star_properties}
\end{figure}

We used two approaches to determine the stellar properties of the host star.
In both cases we assumed an object distance of $94.6\pm0.3$\,pc based on the Gaia DR2 parallax \citep{Gaia2018,Bailer-Jones2018}.
Our first method was based on the X-SHOOTER spectrum and follows the analysis described in \citet{Manara2013}.
We performed a $\chi^2$ fit of the full spectrum using a library of empirical photospheric templates of pre-main sequence stars presented by \citet{Manara2013a,Manara2017}.
The best fit is obtained using the template of the K4 star RXJ1538.6-3916 with an extinction of $A_V=0.0$\,mag.
This converts to an effective temperature of $4590\pm50$\,K and a luminosity of $\log\left(L/L_{\sun}\right)=-0.33\pm0.10$\,dex.
Comparison against isochronal tracks of \citet{Baraffe2015} - hereafter B15 - provides a stellar mass of $1.01\pm 0.08\,M_{\sun}$ and an age of 15 $\pm$ 5\,Myr.
We derived an independent age estimate of the system based on the Lithium-absorption equivalent width of $360\pm20$\,m\AA\;as measured in the X-SHOOTER spectrum.
As presented in panel (a) of Figure~\ref{fig:star_properties}, this provides an age estimate of $17\pm1$\,Myr when compared to the B15 tracks.
The Lithium abundances of the isochrones were converted to Lithium-absorption equivalent widths adopting an initial lithium abundance of $3.28\pm0.05$ \citep{Lodders2009} and using the tables presented in \citet{Soderblom1993}.

An additional check for the stellar properties is by using the photometry.
To constrain the stellar properties of TYC~8998-760-1 we used existing photometry measurements from Tycho-2 \citep{Hog2000}, APASS \citep{Henden2014}, Gaia \citep{Gaia2018}, 2MASS \citep{Cutri2012a}, and WISE \citep{Cutri2012b} catalogues.
Consistent with our previous results, we assumed a negligible extinction and fitted a grid of BT-Settl models \citep{Baraffe2015} with the abundances from \citet{Caffau2011} to the data.
This fit provides an effective temperature of $4573\pm10$\,K and a luminosity of $\log\left(L/L_{\sun}\right)=-0.339\pm0.016$\,dex.
Comparison to the B15 pre-main sequence isochrones plotted in an Hertzsprung-Russell (HR) diagram as presented in panel (b) of Figure~\ref{fig:star_properties}, results in a stellar mass of $1.00\pm0.02\,M_{\sun}$ and a system age of $16.3\pm1.9$\,Myr.

The derived stellar properties for both methods are consistent within their uncertainties.
In Table~\ref{tbl:stellar_properties} we cite the more precise mass, temperature and luminosity estimates for TYC-8998-760-1.
As the determined effective temperature suggests a spectral type of K3 instead of K4 when comparing it to the scale presented in \citet{Pecaut2013}, we adopt the former for our final classification.
For the age of the system, we apply the average of $16.7\pm1.4$\,Myr based on our Lithium-absorption and HR diagram analysis.
This estimate is in good agreement with the average age of LCC of $15\pm3$\,Myr as determined by \citet{Pecaut2016}.

To accurately characterise the companion around TYC~8998-760-1, we determined the magnitudes of the primary in the applied SPHERE and NACO filters.
For all wavelengths shorter than 2500\,nm (i.e. all SPHERE filters) we measured these fluxes directly from  our calibrated X-SHOOTER spectrum.
To assess the stellar magnitudes in $L'$ and $M'$ bands, we used the BT-Settl model instead that we have previously fitted to the available photometric data.
The results of this analysis are presented in Table~\ref{tbl:companion_photometry}.

\subsection{Companion properties}
\label{subsec:results_companion_properties}

We extracted astrometry and magnitude contrasts of the companion for all epochs using the \texttt{SimplexMinimizationModule} of PynPoint as described in \citet{Stolker2019}.
This injects a negative artificial companion into each individual science frame aiming to iteratively minimize the curvature in the final image around the position of the companion using a simplex-based Nelder-Mead algorithm \citep{Nelder1965}.
For the SPHERE data we obtained this template PSF from the non-coronagraphic flux images and
for the NACO data this negative artificial companion was modelled from the unsaturated stellar PSF of the science data itself.
For the latter case we have an individual template for each science frame that directly accounts for the different PSF shapes due to wind effects or varying AO performance.
As the parallactic rotation of the SPHERE datasets is not sufficient to perform ADI-based post-processing strategies, we derotated and median combined the images.
For both NACO datasets, we performed ADI+PCA \citep{PynPoint,Soummer2012} and subtracted one principal component from the images.
We then applied a Gaussian filter with a kernel size equivalent to the pixel scale to smooth pixel to pixel variations before evaluating the curvature in the residual image in an aperture with a radius of one FWHM around the companion.

When studying the residuals after the minimization, it became clear that this analysis method is non-optimal for determining the companion's astrometry and photometry in the SPHERE data.
Whereas in the NACO data the residuals around the companion agree with the average background noise at the same radial separation, the minimisation does not provide similarly smooth results for the SPHERE data.
We attribute this to the different shapes of flux and companion PSFs collected under differing atmospheric conditions.
 
We therefore proceeded with aperture photometry to extract the magnitude contrast of the companion in the SPHERE data and the astrometry was calibrated by a two-dimensional Gaussian fit, instead.
We chose circular apertures with a radius equivalent to the average FWHM measured in the flux images, and used identical apertures around the position of the companion that was determined by the Gaussian fit.
For an accurate estimate of the background noise at this position, we placed several apertures at the same radial separation from the primary.
The average flux within these background apertures was subtracted from the measured flux of the companion.
As a sanity check, we applied this aperture photometry approach also to the NACO data.
The resulting astrometry and photometry of this analysis is consistent with the previously derived values within their uncertainties.

\subsubsection{Astrometric analysis}
\label{subsubsec:astrometric_analysis}

The astrometry of the companion for several epochs and filters is presented in Table~\ref{tbl:companion_astrometry}.
\begin{table}
 \caption{Astrometry of TYC~8998-760-1~b.}
 \label{tbl:companion_astrometry}
 \def\arraystretch{1.2}
\setlength{\tabcolsep}{12pt}
\begin{tabular}{@{}llll@{}}
\hline
Epoch & Filter & Separation & PA\\ 
(yyyy-mm-dd) & & (\arcsec) & (\degr)\\ 
\hline
2017-07-05 & $H$ & $1.715\pm0.004$ & $212.1\pm0.2$ \\
2019-03-17 & $K_s$ & $1.706\pm0.008$ & $212.0\pm0.3$\\
2019-03-23 & $Y2$ & $1.712\pm0.003$ & $212.0\pm0.1$ \\
2019-03-23 & $Y3$ & $1.714\pm0.003$ & $212.0\pm0.1$ \\
2019-03-23 & $J2$ & $1.711\pm0.003$ & $212.0\pm0.1$ \\
2019-03-23 & $J3$ & $1.711\pm0.003$ & $212.0\pm0.1$ \\
2019-03-23 & $H2$ & $1.711\pm0.003$ & $212.0\pm0.1$ \\
2019-03-23 & $H3$ & $1.711\pm0.003$ & $212.0\pm0.1$ \\
2019-03-23 & $K1$ & $1.710\pm0.003$ & $212.0\pm0.1$ \\
2019-03-23 & $K2$ & $1.709\pm0.003$ & $212.0\pm0.1$ \\
2019-05-18 & $L'$ & $1.708\pm0.005$ & $212.6\pm0.2$ \\
2019-06-03 & $M'$ & $1.713\pm0.012$ & $212.4\pm0.4$ \\
\hline
\end{tabular}
\end{table}
As the companion is visible in a single exposure, we extracted its radial separation and position angle directly in the reduced center frames to achieve highest astrometric accuracy.
In these frames we can simultaneously fit the position of the companion and the star behind the coronagraph using the four waffle spots.
We thus do not include the $J$ band measurements in Table~\ref{tbl:companion_astrometry}, as these data were collected without any center frames.

The extracted radial separations and position angles of TYC~8998-760-1~b are mostly consistent within their corresponding uncertainties.
Only in the NACO data we measure a systematically larger position angle compared to the SPHERE astrometry.
This systematic effect has the same magnitude as the applied true north correction of $0\fdg486\pm0\fdg180$ adapted from \citet{MussoBarcucci2019}.
Due to the very consistent SPHERE measurements it is thus likely that this correction factor - which \citet{MussoBarcucci2019} present for reference epochs from 2016 to 2018 - is not valid for our NACO data collected in 2019.
This marginal inconsistency, however, does not affect the further companionship assessments of the object.

\begin{figure}
\centering
\includegraphics[width=\columnwidth]{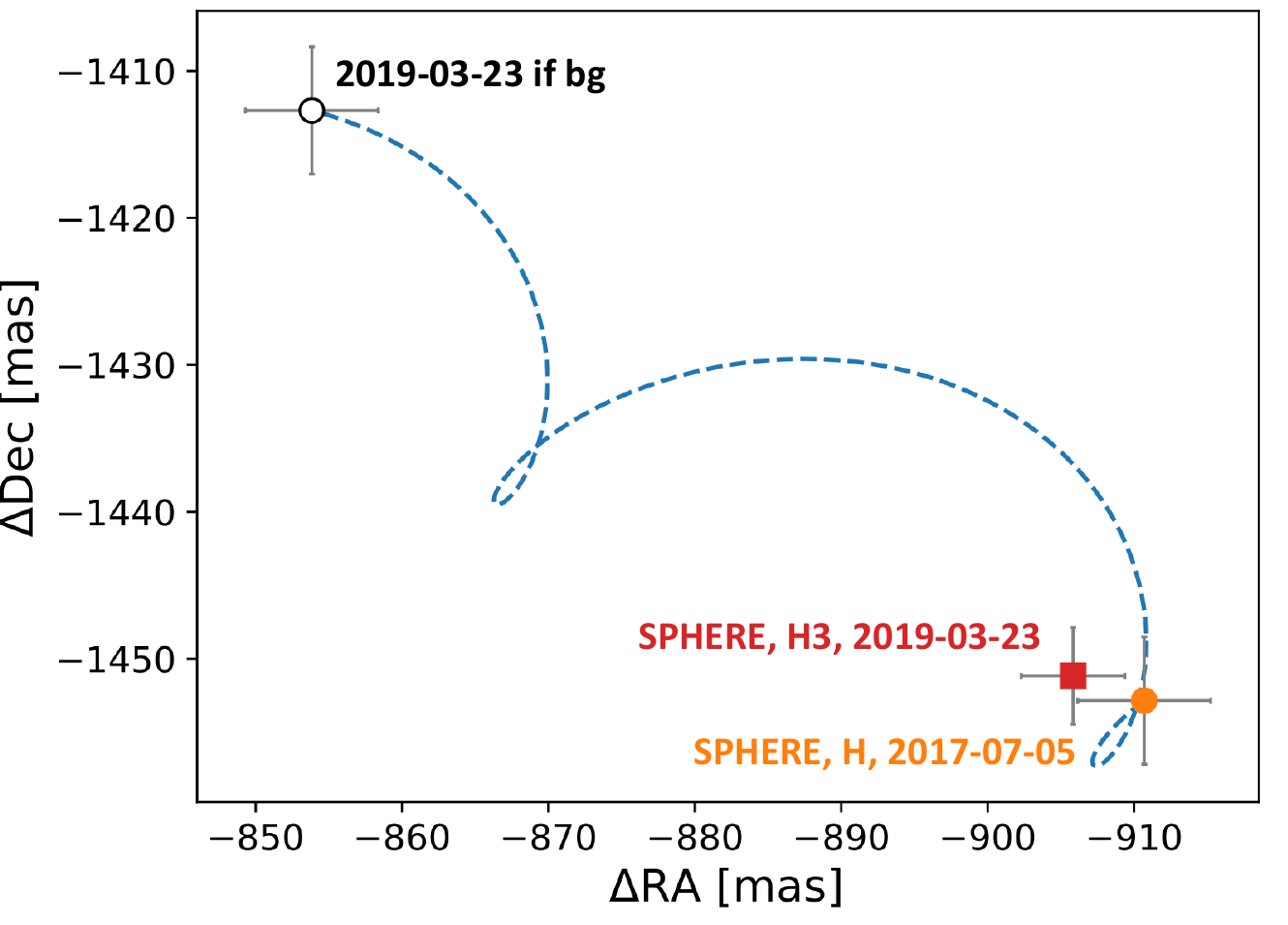}
\caption{
Proper motion plot of the companion south-west of TYC~8998-760-1.
The coordinates are relative offsets to the primary and the blue dashed line represents the trajectory of a static background (bg) object.
}
\label{fig:ppm_analysis_companion}
\end{figure}

Analysis towards common proper motion shows that TYC~8998-760-1~b is clearly co-moving with its host.
As visualised in Figure~\ref{fig:ppm_analysis_companion}, the relative position of the companion is incompatible with a stationary background object at a significance considerably greater than 5$\sigma$.
A similar study was performed for the 15 remaining point sources detected around TYC~8998-760-1.
As presented in Appendix~\ref{sec:proper_motion_analysis_background} their astrometry is highly consistent with background contaminants, instead.

\subsubsection{Photometric analysis}
\label{subsubsec:photometric_analysis}

We present the magnitude contrasts of the companion for all filters in Table~\ref{tbl:companion_photometry}.
\begin{table}
 \caption{Photometry of TYC~8998-760-1~b and its host.}
 \label{tbl:companion_photometry}
 \def\arraystretch{1.2}
\setlength{\tabcolsep}{10pt}
\begin{tabular}{@{}llll@{}}
\hline
Filter & Magnitude star & $\Delta$Mag & Flux companion\\ 
 & (mag) & (mag) & ($\mathrm{erg}\,\mathrm{s}^{-1}\,\mathrm{cm}^{-2}\,\mathrm{\mu m}^{-1}$)\\ 
\hline
\textit{Y2} & 9.47 & $7.56\pm0.21$ & $(0.97\pm0.19)\times10^{-12}$\\
\textit{Y3} & 9.36 & $7.31\pm0.16$ & $(1.13\pm0.16)\times10^{-12}$\\
\textit{J2} & 9.13 & $7.14\pm0.08$ & $(1.16\pm0.08)\times10^{-12}$\\
\textit{J3} & 8.92 & $6.81\pm0.07$ & $(1.37\pm0.08)\times10^{-12}$\\
\textit{H2} & 8.46 & $6.65\pm0.08$ & $(1.04\pm0.07)\times10^{-12}$\\
\textit{H3} & 8.36 & $6.42\pm0.07$ & $(1.12\pm0.07)\times10^{-12}$\\
\textit{K1} & 8.31 & $6.13\pm0.04$ & $(0.77\pm0.03)\times10^{-12}$\\
\textit{K2} & 8.28 & $5.79\pm0.04$ & $(0.88\pm0.03)\times10^{-12}$\\
\textit{J} & 9.02 & $6.71\pm0.38$ & $(1.59\pm0.55)\times10^{-12}$\\
\textit{H} & 8.44 & $7.43\pm0.38$ & $(0.48\pm0.17)\times10^{-12}$\\
$K_s$ & 8.29 & $6.41\pm0.14$ & $(0.54\pm0.07)\times10^{-12}$\\
\textit{L'} & 8.27 & $5.03\pm0.08$ & $(0.26\pm0.02)\times10^{-12}$\\
\textit{M'} & 8.36  & $4.72\pm0.20$ & $(0.16\pm0.03)\times10^{-12}$\\
\hline
\end{tabular}
\end{table}
The SPHERE broadband photometry is rather inconsistent with the dual band measurements, especially in $H$ and $K_s$ band.
This is mainly caused by the very variable observing conditions during these observations.
During the SPHERE $H$ band observations seeing and coherence time between flux and science images degraded from 1\farcs08 to 1\farcs22 and 3.2\,ms to 2.9\,ms, respectively.
In $K_s$ band the conditions were even worse as the seeing increased from 0\farcs74 to 1\farcs11 and the coherence time dropped from 4.5\,ms to 3.5\,ms between flux and science exposures.
Due to these very unstable atmospheric conditions the AO performance was highly variable during these sequences.
Although these fluctuations in flux are included in our statistical uncertainties, the degrading AO performance naturally causes an underestimation of the companion's flux in the science images, leading to an overestimation of the derived magnitude contrast.
Without any additional knowledge of the actual AO performance, it is however not straightforward to correct for this effect.
In our further analysis we thus focus on the results originating from the SPHERE DBI observations that were obtained in more stable weather conditions (see Table~\ref{tbl:observations}).
These variable weather conditions, however, do not affect the astrometric measurements on TYC~8998-760-1~b that we present in Section~\ref{subsubsec:astrometric_analysis}.
As the companion's position angle and separation is directly extracted from the SPHERE center frames, our accuracy is only limited by the precision of the Gaussian fits to the waffle spots and the companion's PSF in these individual frames.

\begin{figure}
\centering
\includegraphics[width=\columnwidth]{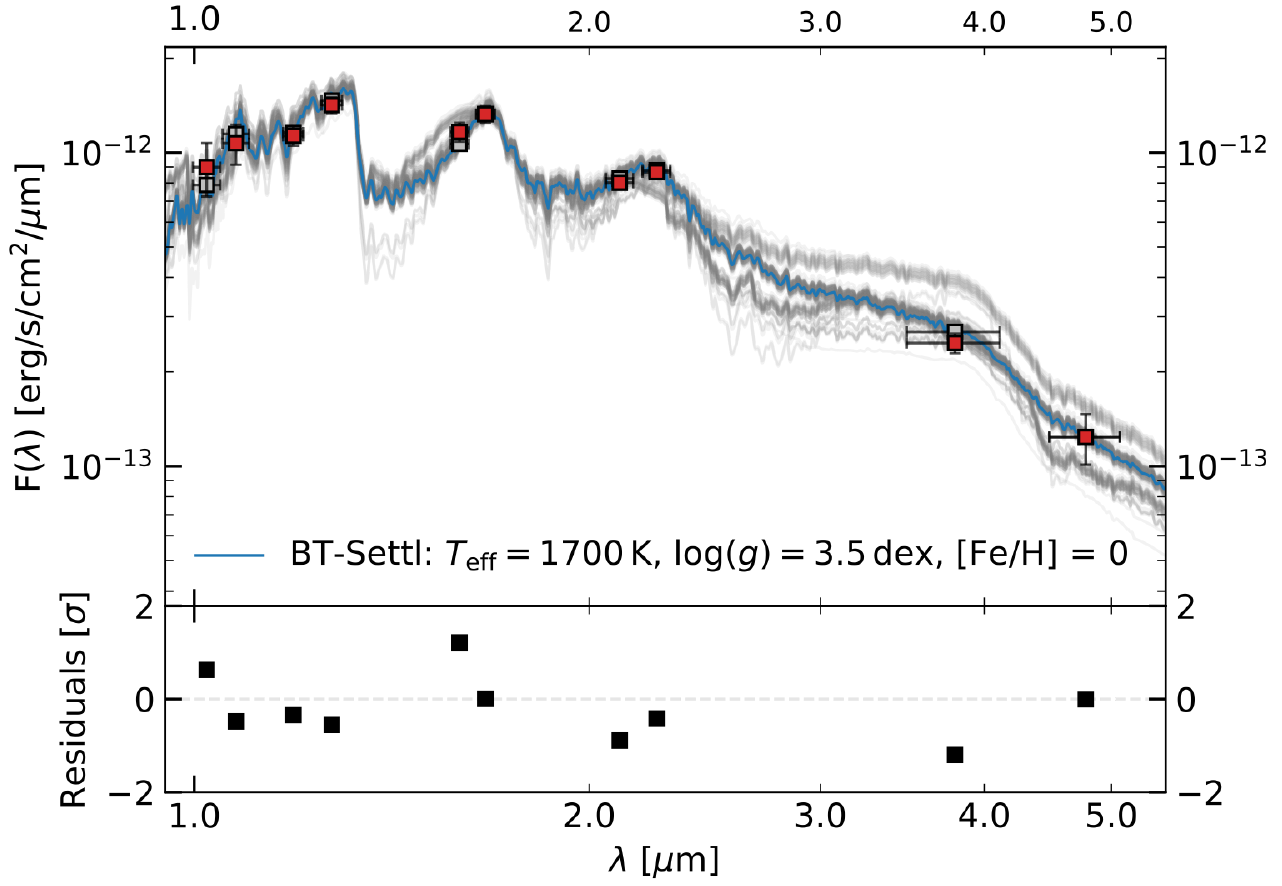}
\caption{
Best-fit result to the spectral energy distribution of TYC-8998-760-1~b.
\emph{Top panel}:
The red squares represent the flux measurements from SPHERE DBI and NACO $L'$ and $M'$ imaging.
The blue line represents the best-fit BT-Settl model \citep{Allard2012} to the data with $T_\mathrm{eff}=1700\,$K, $\log(g)=3.50$\,dex, and solar metallicity and the grey curves represent 200 randomly drawn best-fit models from a Monte Carlo fitting procedure.
The flux of the best-fit model, evaluated at the applied filters, is visualised by the grey squares.
The uncertainties in wavelength direction represent the widths of the corresponding filters.
\emph{Bottom panel}:
Residuals of data and best-fit model.
}
\label{fig:sed_companion}
\end{figure}

To model the companion's SED we converted the apparent magnitudes to physical fluxes using VOSA \citep{Bayo2008}.
These measurements are presented in Table~\ref{tbl:companion_photometry} and visualised as red squares in Figure~\ref{fig:sed_companion}.
To characterise the companion, we fitted a grid of BT-Settl models \citep{Allard2012} to the photometric data by a linear least squares approach.
In agreement with our characterisation of the primary we assumed a negligible extinction and focused on solar metallicity models.
We constrained our input parameter space to effective temperatures between 1200\,K and 2500\,K and surface gravities in the range of 3.0\,dex to 5.5\,dex with step sizes of 100\,K and 0.5\,dex, respectively.
The flux for each model was integrated over the photometric band passes of the applied filters and we determined the scaling that minimises the Euclidean norm of the residual vector.
We compared the resulting residuals for all models from the grid and chose the one that yielded the minimum residual as the best fit.
This is provided by a model with an effective temperature of 1700\,K and a surface gravity of $\log(g)=3.5$\,dex as presented by the blue curve in Figure~\ref{fig:sed_companion}.

\begin{figure*}
\centering
\includegraphics[width=\textwidth]{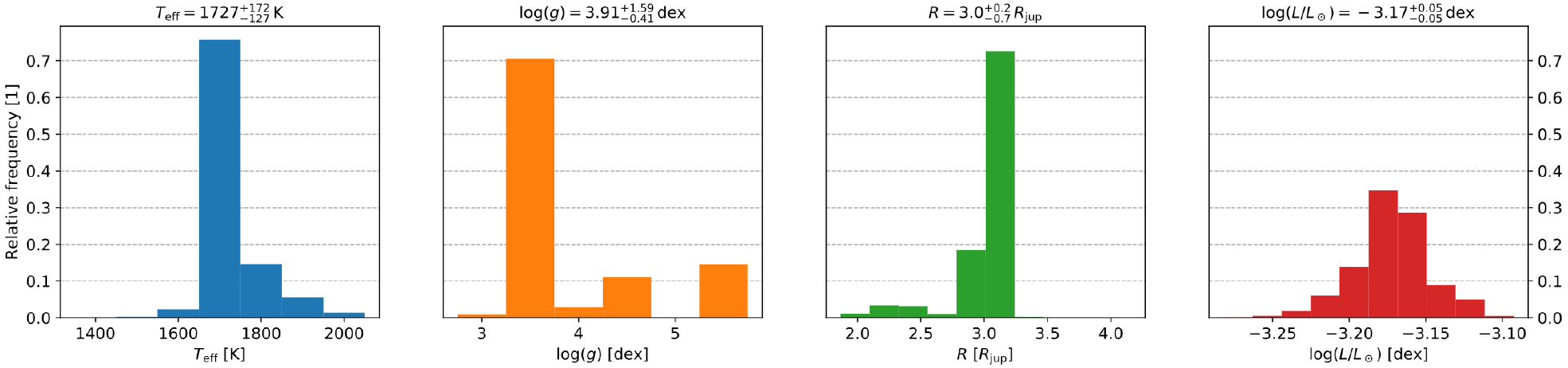}
\caption{
Posterior distributions of best-fit parameters.
The fit is repeated $10^5$ times, drawing each fitted data point from a Gaussian distribution with a standard deviation that is equivalent to the uncertainty.
}
\label{fig:sed_fit_posterior}
\end{figure*}

To evaluate the the impact of the photometric uncertainties on the resulting best fit model, we repeated the fitting procedure $10^5$ times, drawing the fitted fluxes from a Gaussian distribution centered around the actual data point and using the uncertainty as standard deviation of the sampling.
In Figure~\ref{fig:sed_companion}, we show 200 randomly selected best fit models from this Monte Carlo approach as indicated by the grey curves.
The posterior distributions for the best-fit parameters are presented in Figure~\ref{fig:sed_fit_posterior}.
This procedure provides estimates of $T_\mathrm{eff}=1727^{+172}_{-127}\,$K, $\log\left(g\right)=3.91^{+1.59}_{-0.41}$, $R=3.0^{+0.2}_{-0.7}\,R_\mathrm{jup}$, and $\log\left(L/L_{\sun}\right)=-3.17^{+0.05}_{-0.05}\,$dex for the companion's effective temperature, surface gravity, radius, and luminosity, respectively.
The uncertainties of these values are determined as the 2.5 and 97.5 percentiles of the corresponding posterior distributions.
Both radius and luminosity depend on the distance to the system, which is constrained by Gaia DR2 astrometry.
The radius estimate arises from the scaling factor that needs to be applied to the model and the luminosity is obtained by integrating the resulting model over the entire wavelength range.
We note that the predicted radius is larger than the usual value of  $\sim1\,R_\mathrm{jup}$ that is associated with gas giant planets and brown dwarfs \citep[e.g.][]{Chabrier2009}.
This unexpected property is discussed in Section~\ref{subsec:discussion_companion_properties}.

\subsubsection{Companion mass}
\label{subsubsec:analysis_companion_mass}

To convert the derived photometric properties of the companion to a mass, we used BT-Settl isochrones \citep{Allard2012} that we evaluated at the derived system age of 16.7$\pm$1.4\,Myr.
As we only fitted photometric data that does not resolve any lines or molecular features, the object's surface gravity is not strongly constrained from our analysis.
We base our mass estimate on the better constrained effective temperature and luminosity of the companion instead.
Comparing these values to BT-Settl isochrones yields masses of $12.1^{+1.7}_{-1.6}\,M_\mathrm{jup}$ and $15.7^{+1.0}_{-0.4}\,M_\mathrm{jup}$ for measured temperature and luminosity, respectively.
We obtained similar mass estimates when using the AMES-dusty isochrones  \citep{Allard2001,Chabrier2000} instead of the the BT-Settl models.

To test these results, we converted the absolute magnitudes of the companion to mass estimates using the BT-Settl isochones evaluated at the corresponding band passes\footnote{The models were downloaded from \url{http://perso.ens-lyon.fr/france.allard/}.}.
For the SPHERE data this gives values consistent with our previous mass estimates in the range of 14\,$M_\mathrm{jup}$ to 16\,$M_\mathrm{jup}$. 
In the thermal infrared we obtain masses of approximately 18\,$M_\mathrm{jup}$ and 25\,$M_\mathrm{jup}$ for the absolute $L'$ and $M'$ magnitudes.
This gradient towards longer wavelengths is usual for sub-stellar companions, as these are often redder than the predictions from the models \citep{Janson2019}.

We additionally determined the spectral type of the companion following the analysis demonstrated in \citet{Janson2019}.
This analysis was performed analogously to the SED fit described before;
it was however confined to the SPHERE photometry, because the input models only support this wavelength coverage.
Using the empirical spectra for M-L dwarfs of \citet{Luhman2017} we derive a best-fit spectral type of L0.
This is equivalent to the spectral type derived for HIP~79098~(AB)b \citep{Janson2019}, which is indeed an ideal object for comparison, as it is also located in Sco-Cen -- though in the Upper Scorpius sub-group instead of LCC -- with an estimated age of $10\pm3$\,Myr.
The absolute magnitudes for the companion around TYC~8998-760-1 are approximately 1.5\,mag fainter than the values derived for HIP~79098~(AB)b, supporting the theory that TYC~8998-760-1~b is less massive than the object of this comparison, for which \citet{Janson2019} derive a mass range of $16-25\,M_\mathrm{jup}$.

\begin{figure}
\centering
\includegraphics[width=\columnwidth]{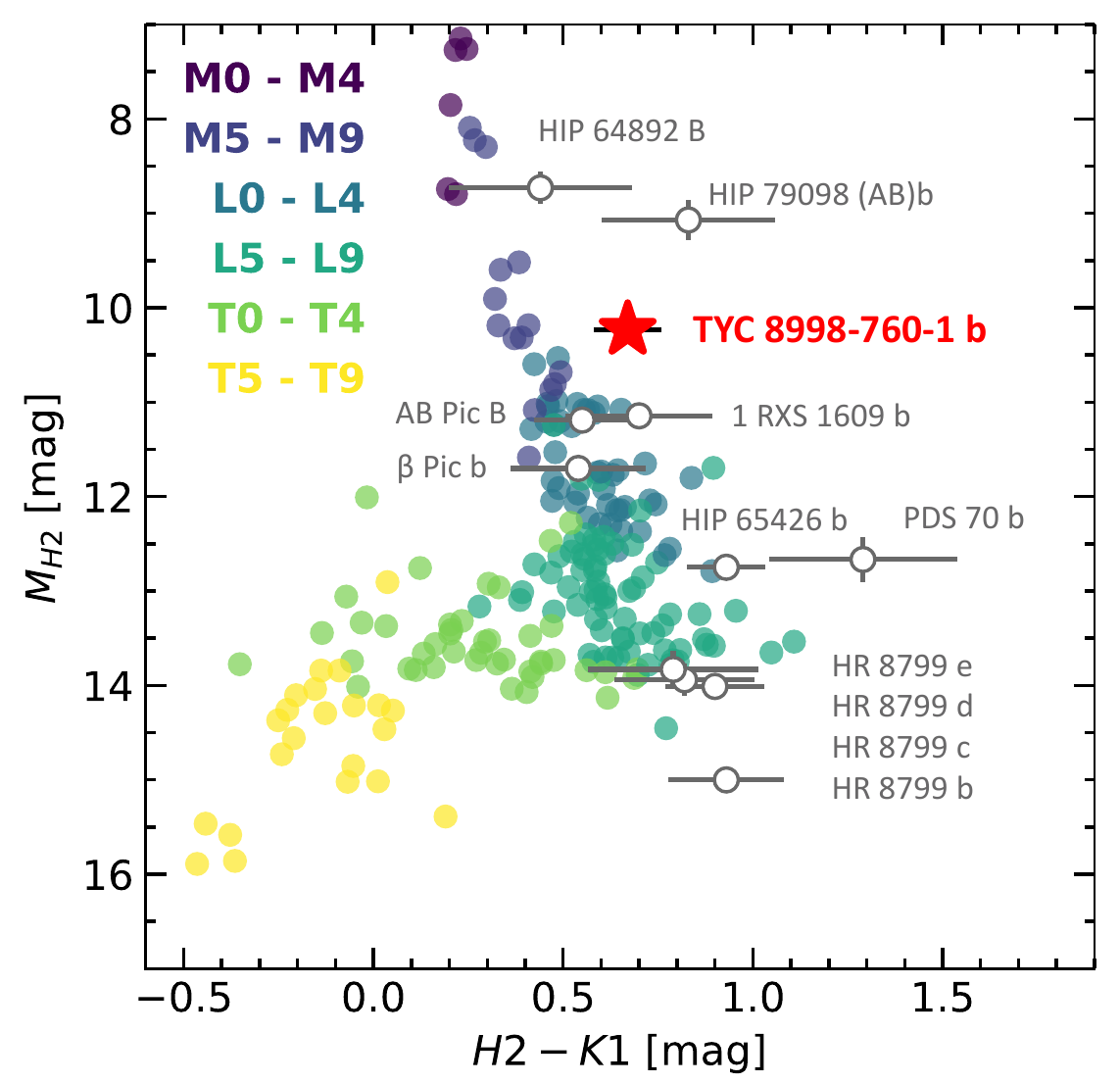}
\caption{
Colour-magnitude diagram for TYC~8998-760-1~b.
The filled circles indicate the colour-magnitude evolution of M, L and T field dwarfs, whereas the white markers indicate companions that were directly imaged around young stars.
TYC~8998-760-1~b - highlighted by the red star - is located at the transition stage between late M and early L dwarfs and is considerably redder than the corresponding evolved counterparts of similar spectral type.
}
\label{fig:cmd_companion}
\end{figure}

To verify the derived properties, we compared the colour of TYC~8998-760-1~b to that of known sub-stellar companions of similar spectral type.
Based on the NIRSPEC Brown Dwarf Spectroscopic Survey \citep{McLean2003,McLean2007}, the IRTF Spectral library \citep{Rayner2009,Cushing2005}, and the L and T dwarf data archive \citep{Knapp2004,Golimowski2004,Chiu2006}, we compiled a sample of M, L, and T dwarfs.
The spectra of these objects were evaluated at the bandpasses of the SPHERE $H2$ and $K1$ filters that we chose for the colour analysis.
To determine the absolute magnitudes of these field dwarfs we used distance measurements provided by Gaia DR2 \citep{Gaia2018,Bailer-Jones2018}, the Brown Dwarf Kinematics Project \citep{Faherty2009}, and the Pan-STARRS1 3$\pi$ Survey \citep{Best2018}.
Targets without any parallax measurement were discarded from the sample.
In addition to these field objects, we compared the colour of TYC~8998-760-1~b to photometric measurements\footnote{For companions that have not been observed with the identical combination of SPHERE $H2$ and $K1$ dual band filters, we based the presented magnitudes and colours on the corresponding broadband photometry, instead.} of confirmed sub-stellar companions \citep[based on data from][]{Cheetham2019,Janson2019,Lafreniere2008,Chauvin2005,Currie2013,Bonnefoy2011,Keppler2018,Muller2018,Chauvin2017a,Zurlo2016}.
The results of this analysis are presented in a colour-magnitude diagram in Figure~\ref{fig:cmd_companion}.
TYC~8998-760-1~b is located at the transition between late M and early L-type dwarfs, which is in very good agreement with the previously assigned spectral type of L0.
As observed for many other young, directly imaged L-type companions, TYC~8998-760-1~b is considerably redder than the sequence of evolved field dwarfs of similar spectral type.
This appearance is associated with lower surface gravities of these young objects in comparison to their field counterparts \citep[e.g.][]{Gizis2015, Janson2019}.

All our analyses, therefore, indicate that the detected companion is sub-stellar in nature.
Accounting for the spread among the various methods used to infer the object's mass, we adopt a conservative estimate of $14\pm3\,M_\mathrm{jup}$, yielding a mass ratio of $q=0.013\pm0.003$ between primary and companion.
We conclude that TYC-8998-760-1~forb is a sub-stellar companion to TYC-8998-760-1 at the boundary between giant planets and low mass brown dwarfs.
Further studies at higher spectral resolution are required to confine this parameter space and to test the planetary nature of the object.

\subsection{Detection limits}
\label{subsec:detection_limits}

\begin{figure}
\centering
\includegraphics[width=\columnwidth]{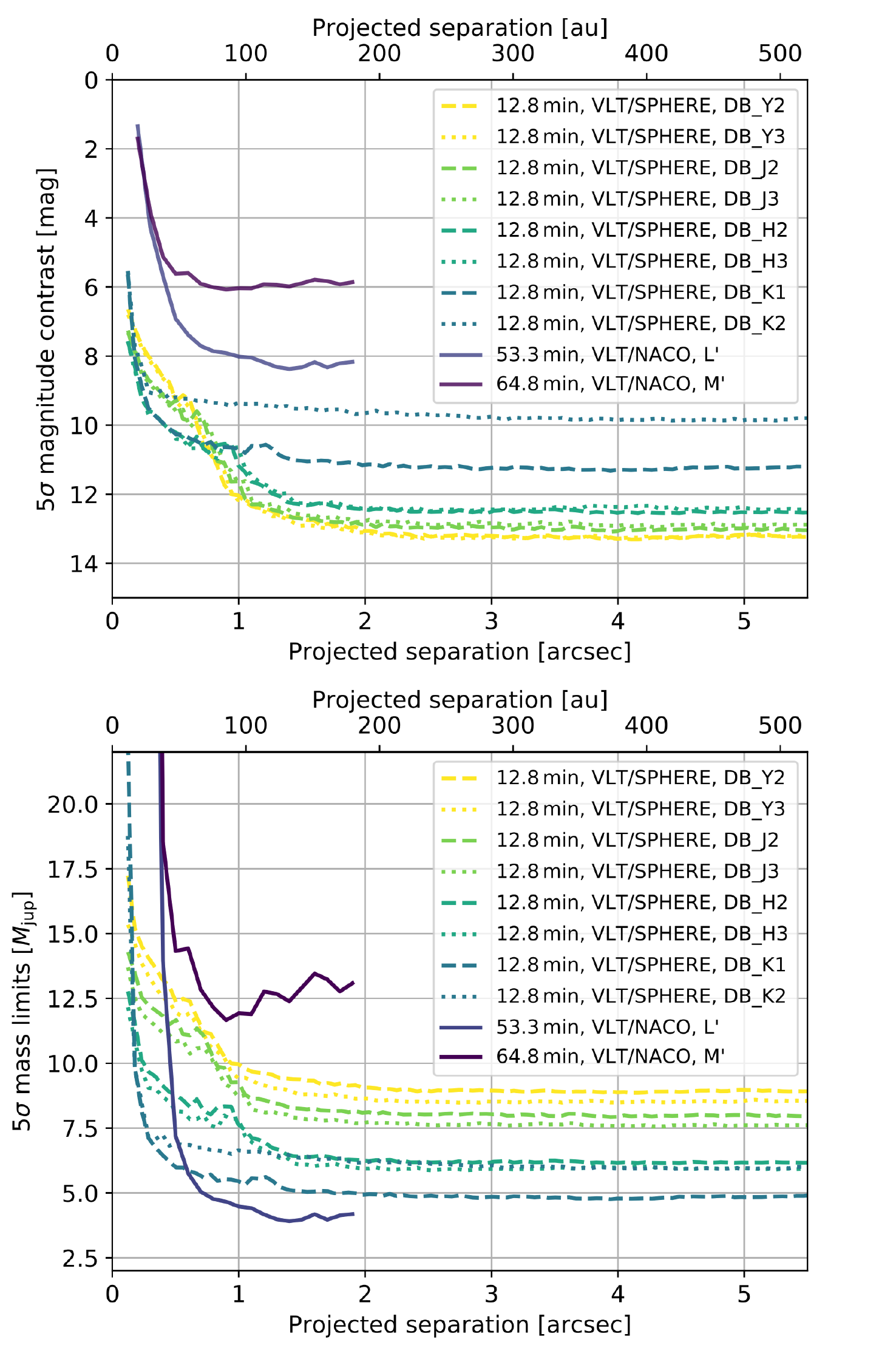}
\caption{
Detection limits for SPHERE/DBI and NACO datasets.
\emph{Upper panel}: 
Magnitude contrast as a function of angular separation.
\emph{Lower panel}:
Mass limits as a function of angular separation.
The magnitude contrast is converted to masses via AMES-dusty \citep{Allard2001,Chabrier2000} models.
}
\label{fig:detection_limits}
\end{figure}

To assess our sensitivity  to further companions in the system, we determined the contrast limits for each of the datasets.
For the SPHERE data, which do not provide a large amount of parallactic rotation, we did not perform any PSF subtraction.
Instead we determined the contrast in the derotated and median combined images by measuring the standard deviation of the residual flux in concentric annuli around the star.
To exclude flux of candidate companions that might distort these noise measurements, we performed a 3$\sigma$ clipping of the flux values inside the annuli, before calculating the standard deviation of the remaining pixels.
The annuli have widths of the FWHM at the corresponding wavelength and we evaluate the statistics at radial separations between 0\farcs1 and 5\farcs5 with a step size of 50\,mas.
With these noise terms and the peak flux of the PSF in the corresponding median flux image, we derived the 5$\sigma$ contrast curves for the SPHERE data, presented in the top panel of Figure~\ref{fig:detection_limits}.
Due to the poor weather conditions and shorter integration times, we neglect the SPHERE broadband imaging data for this analysis.

The NACO data was analysed with the \texttt{ContrastCurveModule} of PynPoint.
For both $L'$ and $M'$ data we injected artificial planets into the data and fitted one principal component for PSF subtraction before de-rotation.
The planets were injected at six equidistantly distributed angles with radial separations increasing from 0\farcs2 to 2\farcs0 and a step size of 100\,mas.
The magnitude of the injected planets was optimised so that these are detected at 5$\sigma$ significance applying an additional correction for small sample statistics at small angular separations \citep{Mawet2014}.
To obtain the final contrast curves as presented in the top panel of Figure~\ref{fig:detection_limits} we averaged the data along the azimuthal dimension.

To convert the derived magnitude contrasts to detectable planetary masses we used the AMES-dusty models \citep{Allard2001,Chabrier2000} and evaluated the isochrones at a system age of 16.7\,Myr.
The SPHERE observations provide the best performance for small angular separations.
The $H2$ data rules out any additional companions more massive than $12\,M_\mathrm{jup}$ for separations larger than 120\,mas.
This is equivalent to ruling out additional stellar or brown dwarf companions separated farther than 12\,au from TYC~8998-760-1.
For angular separations larger than 0\farcs5 up to approximately 2\arcsec, NACO $L'$ band imaging yields the tightest constraints for additional companions in the system.
For separations in the range of 1\arcsec to 2\arcsec we can rule out additional companions that are more massive than approximately $4\,M_\mathrm{jup}$.
Farther out, the $H2$ background limit is approximately $5\,M_\mathrm{jup}$.

Due to deeper integrations in the SPHERE observations collected on the night of March 23, 2019, we detect additional point sources to the 16 objects that were found in the first epoch data from July 5, 2017.
The contrasts of these objects are above the derived detection limits.
Statistical evaluation based on the first epochs already indicates a very high fraction of background contaminants in the IRDIS field of view around TYC~8998-760-1;
as we do not have additional data to test the proper motion of these new candidate companions we cannot entirely rule out the possibility that these are co-moving with TYC~8998-760-1.

\section{Discussion}
\label{sec:discussion}

\subsection{Companion properties}
\label{subsec:discussion_companion_properties}

Whilst effective temperature, surface gravity and luminosity of TYC~8998-7601~b that we have derived in Section~\ref{subsubsec:photometric_analysis} seem to agree with general properties of similar low-mass companions \citep[e.g.][]{Bonnefoy2013,Chauvin2017a} the radius estimate of $R=3.0^{+0.2}_{-0.7}\,R_\mathrm{jup}$ is larger than expected from these analogous systems.
Empirical data suggest an almost constant radius of approximately 1\,$R_\mathrm{jup}$ for planets in the range of $1\,M_\mathrm{jup}$ up to stellar masses \citep[e.g.][]{Chabrier2009} - but these relations are derived from field populations of sub-stellar objects.
Their young, gravitationally bound counterparts tend to be inflated instead as these are still contracting \citep{Baraffe2015}.
This leads to earlier spectral types, lower surface gravities, and larger radii of young companions in comparison to field objects of the same mass \citep{Asensio-Torres2019}.
Furthermore, the constraints that are imposed on the radius are only very weak.
The lower bound from the Monte Carlo analysis already implies that smaller radii are not ruled out by our best-fit models.
As the masses that are derived from effective temperature, luminosity, individual photometry, and spectral type are all in very good agreement, it is unlikely that the object is not a low-mass companion to TYC~8998-760-1.

Another possible explanation for the radius anomaly might be given by the scenario that TYC~8998-760-1~b is an unresolved binary with two components of near equal brightness.
To test this hypothesis, we repeated the SED modeling, allowing for two objects contributing to the observed photometry.
The best-fit result is obtained by binary components with effective temperatures of 1700\,K and 1800\,K and corresponding radii of $1.6\,R_\mathrm{jup}$ and $2.1\,R_\mathrm{jup}$.
These results are in better agreement with potential radii of inflated, young sub-stellar objects \citep{Baraffe2015}.
As the PSF of TYC~8998-760-1~b is azimuthally symmetric, this potential binary pair of nearly equal brightness would have to be unresolved in our data.
Applying the FWHM for our observations at highest angular resolution in Y2 band (see Table~\ref{tbl:observations}) implies that a binary companion must have a angular separation smaller than 37.2\,mas to be unresolved in the data.
At the distance of this system this translates to a physical separation smaller than 3.5\,au, which lies well within the Hill sphere of a secondary with a mass of approximately $14\,M_\mathrm{jup}$. 
Although this hypothesis might explain the large radius that we find for TYC~8998-760-1~b, additional data of the companion is required to thoroughly test this scenario of binarity.
An infra-red medium resolution spectrum of the companion would thus be very valuable for confirming this hypothesis.

\subsection{Comparison to other directly imaged sub-stellar companions}
\label{subsec:comparison_directly_images_planets}

\begin{figure}
\centering
\includegraphics[width=\columnwidth]{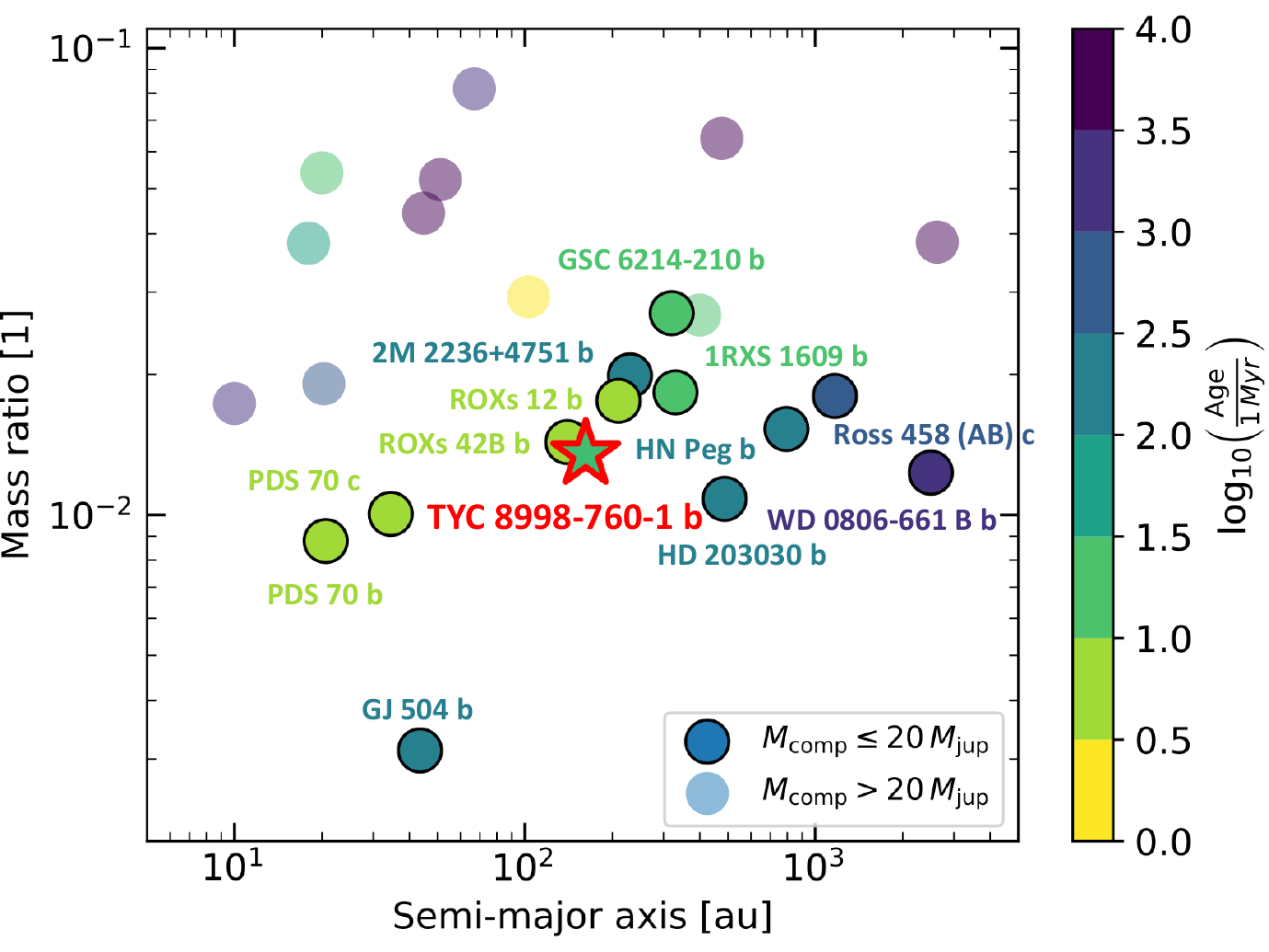}
\caption{
Directly imaged sub-stellar companions around solar-mass stars.
For the sample selection we chose host stars with masses in the range of $0.6\,M_{\sun}$ and $1.4\,M_{\sun}$
We present the mass ratio $q$ between companion and primary as a function of radial separation to the host.
The colour indicates the age of the corresponding system.
}
\label{fig:comparison_other_companions}
\end{figure}

Although tens of low-mass, sub-stellar companions have been directly imaged, the majority of the host stars are either more massive than the Sun \citep[e.g.][]{Lagrange2010,Marois2008,Rameau2013,Chauvin2017a,Carson2013,Janson2019}, are located at the lower end of the stellar mass distribution \citep[e.g.][]{Luhman2005,Delorme2013,Artigau2015,Bejar2008,Luhman2009,Rebolo1998,Kraus2014,Bowler2013,Gauza2015,Naud2014,Itoh2005}, or of sub-stellar nature themselves \citep[e.g.][]{Todorov2010,Gelino2011,Liu2012}.
The sample of planetary mass companions that are unambiguously confirmed around solar-type stars is still small, containing PDS~70~b and c \citep{Keppler2018,Haffert2019}, 2M~2236+4751~b \citep{Bowler2017}, AB~Pic~b \citep{Chauvin2005}, 1RXS~1609~b \citep{Lafreniere2008}, HN~Peg~b \citep{Luhman2007}, CT~Cha~b \citep{Schmidt2008}, HD~203030~b \citet{Metchev2006}, and GJ~504~b \citet{Kuzuhara2013}.
This selection was compiled\footnote{For this analysis we used the \url{http://exoplanet.eu/} database \citep{Schneider2011}} applying conservative mass thresholds in the range of $0.6\,M_{\sun}$ to $1.4\,M_{\sun}$ for host stars to be considered solar type.
In Fig.~\ref{fig:comparison_other_companions}, we visualise the properties of TYC~8998-760-1~b among this sample of directly imaged sub-stellar companions around solar-mass stars.
To estimate the semi-major axis of the object, we use the projected separation of 162\,au that we derived earlier.
This value is thus a lower limit of the actual semi-major axis, as it is the case for many directly imaged companions on wide orbits.

From Fig.~\ref{fig:comparison_other_companions} it is apparent that TYC~8998-760-1 is among the youngest systems with a directly imaged sub-stellar companion around a solar-mass host star.
Its mass ratio $q$ is one of the smallest within the sample, only surpassed by HD~203030~b, GJ~504~b, and both planets around PDS~70.
The distance at which it is detected is interesting as it is well separated from the host.
This facilitates long-term monitoring and spectroscopic characterisation of the companion with both ground and space based missions.
Near infrared observations towards the photometric variability of the object would help to constrain its rotation period and potential cloud coverage \citep[e.g][]{Yang2016};
additional spectroscopic data will allow to constrain the mass of TYC~8998-760-1~b and to determine molecular abundances in its atmosphere \citep[e.g.][]{Hoeijmakers2018}.

\subsection{Formation scenarios}
\label{subsec:discussion_formation_scenarios}

The origin of giant planetary-mass companions at large separations from their host stars is a highly debated topic.
Studies by \citet{Kroupa2001} and \citet{Chabrier2003} argue that these objects can form in situ and represent the lower mass limit of multiple star formation via fragmentation processes in the collapsing protostellar cloud. 
If the companion has formed via the core accretion channel \citep{Pollack1996,Alibert2005,DodsonRobinson2009,Lambrechts2012} or via gravitational instabilities of the protoplanetary disc \citep{Boss1997,Rafikov2005,Durisen2007,Kratter2010,Boss2011} this must have happened closer to the star and after formation, the protoplanet needs to be scattered to the large separation at which it is observed.
For regions with a high number density of stars such as Sco-Cen, also capture of another low-mass member of the association needs to be considered as a potential pathway of producing wide orbit companions \citep[e.g.][]{Varvoglis2012,Goulinski2018}.
TYC~8998-760-1~b is an ideal candidate to test potential scenarios of (i) formation closer to the host and scattering to its current location, (ii) in-situ formation, and (iii) capture of a low mass Sco-Cen member.

Scenario (i) requires a third component in the system in addition to host star and companion.
This component has to be more massive than the companion to scatter the protoplanet off the system to its current location.
Even though the detection limits of our high-contrast observations rule out additional companions that are more massive than $12\,M_\mathrm{jup}$ for projected separations that are larger than 12\,au, this does not rule out a binary companion in a close orbit around TYC~8998-760-1.
To constrain the parameter space of a close, massive companion in the system, reflex motion measurements of the host star are required.
This analysis could be performed by combining our high-contrast imaging data with additional radial velocity observations of the system as for instance presented by \citet{Boehle2019}.
High-precision astrometry provided by future data releases of the Gaia mission \citep{Gaia2016} will be valuable to identifying potential close-in binaries.

One way to discriminate between the three potential formation scenarios is provided by a precise determination of TYC~8998-760-1~b's orbit.
This can be achieved by monitoring of the relative astrometric offset between primary and secondary in combination with additional radial velocity measurements.
The primary's radial velocity is measured by Gaia as $12.8\pm1.4\,\mathrm{km}\mathrm{s}^{-1}$ and for the companion - as it is reasonably far separated from the host - this will be accessible by medium resolution spectroscopy.
Polarimetric observations of the target and detection of a potential circumstellar or even circumplanetary disc around either of the components would impose further constraints on the orbital dynamics of the system.

With the currently available data it is not possible to unambiguously identify the mechanism that shaped the appearance of the young solar system around TYC~8998-760-1, but with future observations as outlined in the previous paragraphs, it should be possible to discern which is the most likely scenario that shaped the architecture of this young, solar-like system.

\section{Conclusion}
\label{sec:conclusions}

After the discovery of a shadowed protoplanetary disc at transition stage around Wray~15-788 \citep{Bohn2019}, we report the detection of a first planetary mass companion within the scope of YSES.
The companion is found around the K3IV star TYC~8998-760-1, located in the LCC subgroup of Sco-Cen.
Using X-SHOOTER and archival photometric data, we determine a mass of $1.00\pm0.02\,M_{\sun}$, an effective temperature of $4573\pm10\,$K, a luminosity of $\log\left(L/L_{\sun}\right)=-0.339\pm0.016\,$dex, and an age of $16.7\pm1.4$\,Myr for the primary. 
The companion is detected at a projected separation of approximately 1\farcs7 which translates to a projected physical separation of 162\,au at the distance of the system.
Fitting the companion's photometry with BT-Settl models provides an effective temperature of $T_\mathrm{eff}=1727^{+172}_{-127}\,$K, a surface gravity of $\log\left(g\right)=3.91^{+1.59}_{-0.41}$, a radius of $R=3.0^{+0.2}_{-0.7}\,R_\mathrm{jup}$, and a luminosity of $\log\left(L/L_{\sun}\right)=-3.17^{+0.05}_{-0.05}\,$dex.
At the age of the system we adopt a mass estimate of $14\pm3\,M_\mathrm{jup}$, which is equivalent to a mass ratio of 
$q=0.013\pm0.03$ between primary and secondary.
TYC~8998-760-1~b is among the youngest and least massive companions that are directly detected around solar-type stars.
The large radius we have derived suggests that the companion is either inflated, or is an unresolved binary in a spatially unresolved orbit with a semi-major axis smaller than $3.5\,$au.
From our high-contrast imaging data we can exclude any additional companions in the system with masses larger than $12\,M_\mathrm{jup}$ at separations larger than 12\,au.
This discovery opens many pathways for future ground and space-based characterisation of this solar-like environment at a very early stage of its evolution.

\section*{Acknowledgements}

We thank the anonymous referee for the valuable feedback that helped improving the quality of the manuscript.

The research of AJB and FS leading to these results has received funding from the European Research Council under ERC Starting Grant agreement 678194 (FALCONER).

Part of this research was carried out at the Jet Propulsion Laboratory, California Institute of Technology, under a contract with the National Aeronautics and Space Administration.

The research leading to these results has received funding from the European Research Council (ERC) under the EuropeanUnion's Horizon 2020 research and innovation programme (grant agreement no. 679633; Exo-Atmos).

CFM acknowledges an ESO fellowship.
This project has received funding from the European Union's Horizon 2020 research and innovation programme under the Marie Sklodowska-Curie grant agreement No 823823 (DUSTBUSTERS).
This work was partly supported by the Deutsche Forschungs-Gemeinschaft (DFG, German Research Foundation) - Ref no. FOR 2634/1 TE 1024/1-1.

This research has used the SIMBAD database, operated at CDS, Strasbourg, France \citep{Wenger2000}. This work has used data from the European Space Agency (ESA) mission
{\it Gaia} (\url{https://www.cosmos.esa.int/gaia}), processed by the {\it Gaia}
Data Processing and Analysis Consortium (DPAC,
\url{https://www.cosmos.esa.int/web/gaia/dpac/consortium}). Funding for the DPAC
has been provided by national institutions, in particular the institutions
participating in the {\it Gaia} Multilateral Agreement.

This publication makes use of VOSA, developed under the Spanish Virtual Observatory project supported by the Spanish MINECO through grant AyA2017-84089. 

We used the \emph{Python} programming language\footnote{Python Software Foundation, \url{https://www.python.org/}}, especially the \emph{SciPy} \citep{SciPy}, \emph{NumPy} \citep{numpy}, \emph{Matplotlib} \citep{Matplotlib}, \emph{scikit-image} \citep{scikit-image}, \emph{scikit-learn} \citep{scikit-learn}, \emph{photutils} \citep{photutils}, and \emph{astropy} \citep{astropy_1,astropy_2} packages.
We thanks the writers of these software packages for making their work available to the astronomical community.




\bibliographystyle{mnras}
\bibliography{mybib} 




\newpage
\appendix

\section{Proper motion analysis of other point sources}
\label{sec:proper_motion_analysis_background}

\begin{figure*}
\centering
\includegraphics[width=\textwidth]{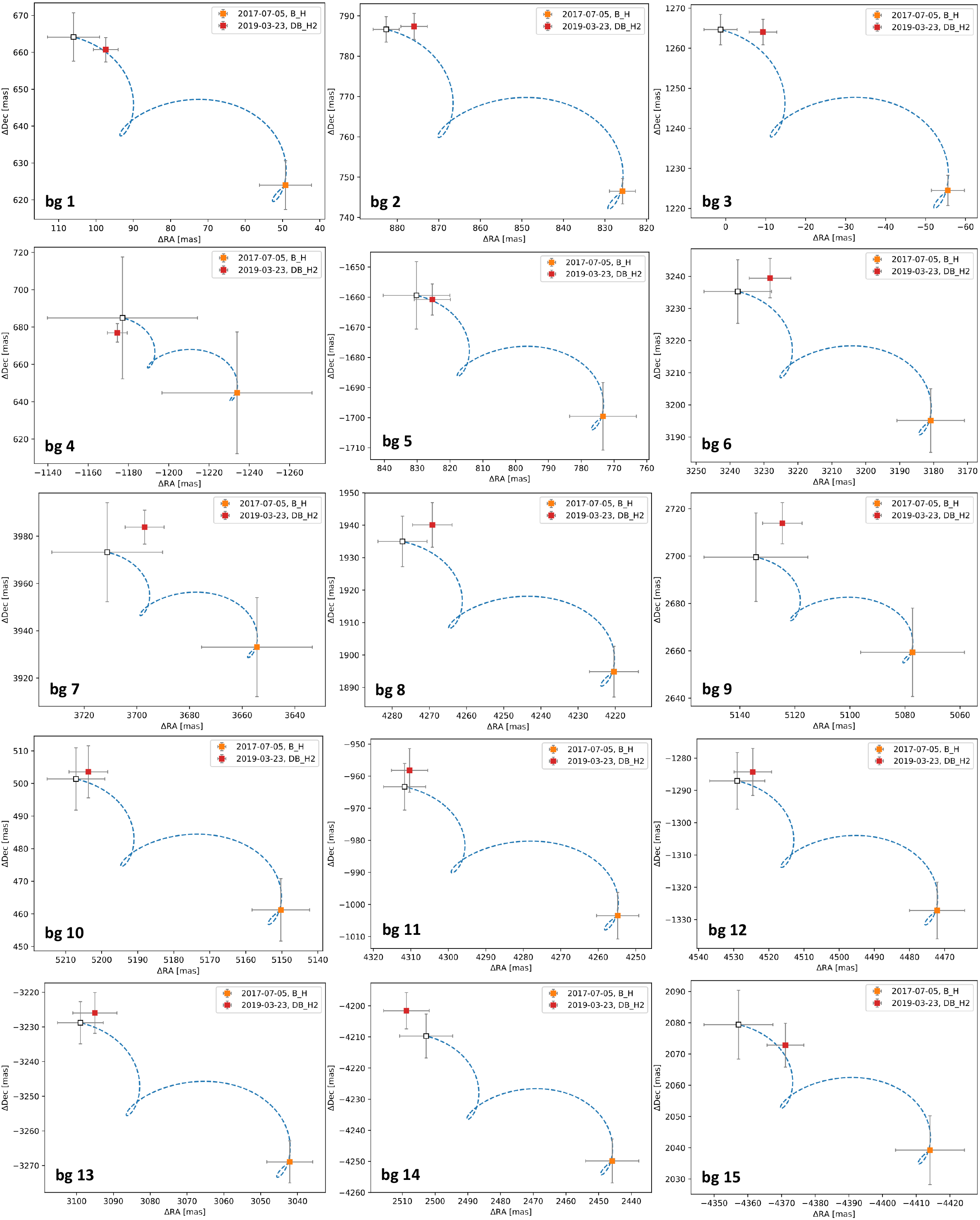}
\caption{
Proper motion analysis of other candidate companions around TYC~8998-760-1.
The coordinates are relative offsets to the primary and the blue dashed line represents the trajectory of a static background (bg) object.
The white marker along that trajectory indicates the expected relative position of a static background object for the second epoch data.
}
\label{fig:ppm_analysis_background}
\end{figure*}

In our first epoch data, we detect 16 point sources around TYC~8998-760-1.
All these candidate companions are re-detected in our deeper second epoch data from March 23, 2019.
We analysed the relative motion of all these object towards common proper motion with the primary.
As presented in Figure~\ref{fig:ppm_analysis_background} all candidate companions but TYC~8998-760-1~b have to be considered background contaminants, as their relative positions are not compatible with a bound companion.
In most cases our measurements agree well with the predicted trajectory of a static background object.
Small deviations from this prediction indicate an intrinsic non-zero proper motion of the object, instead.


\bsp	
\label{lastpage}
\end{document}